\newcommand{\snn} {\mbox{$\sqrt{s_{NN}}$}}

\def\Journal#1#2#3#4{#1 {\bf #2}, #3 (#4)}
\def\JournalAPPEAR#1#2{#1 {\bf #2}}

\def\NPA{Nucl. Phys.~A}

\def\PRC{Phys. Rev.~C}

\documentclass[12pt]{iopart}
\usepackage{epsf}
\usepackage{graphicx}

\begin{document}

\title[]{Hadron Production at Intermediate $p_T$ at RHIC}

\author{Tatsuya Chujo
\footnote[3]{tatsuya.chujo@vanderbilt.edu}
for the PHENIX Collaboration\footnote{See Appendix for the full collaboration list.}
}

\address{Vanderbilt University, Nashville, TN 37235, USA}

\begin{abstract}
Large proton and antiproton enhancement with respect to pions has
been observed at intermediate transverse momentum $p_T \approx$ 2-5 GeV/$c$ 
in Au+Au collisions at RHIC. To investigate the possible source of this 
anomaly, the production of $\phi$ mesons and two particle angular 
correlations triggered by mid-$p_T$ baryons or mesons are studied. 
We also present the first measurement of proton and antiproton production 
at $\snn = 62.4$ GeV in Au+Au collisions, which aims to study the energy 
dependence of the observed baryon enhancement.
\end{abstract}


\section{Introduction -- Baryon Anomaly at RHIC}

One of the most striking observations in the central heavy ion collisions 
at the Relativistic Heavy Ion Collider (RHIC) is a large enhancement of 
baryons and antibaryons relative to pions at intermediate $p_{T} \approx$ 2--5 
GeV/$c$~\cite{ppg015,ppg026}, while the neutral pions~\cite{ppg014} and inclusive 
charged hadrons~\cite{ppg023} are strongly suppressed at those $p_T$. The (anti)proton 
to pion ratio is enhanced by almost a factor of three when one compares peripheral 
reactions to the most central Au+Au reactions. In this region of $p_{T}$ fragmentation 
dominates the particle production in $pp$ collisions. It is expected that fragmentation is 
independent of the colliding system - hence the large baryon fraction observed at RHIC 
comes as a surprise.     

Large proton to pion ratios have also been observed in heavy ion collisions 
at the AGS and SPS. However, the  large net baryon density at these energies indicates 
that most of the protons come from the initial state of the reaction. The anti-proton 
yields at the SPS and AGS are one and three orders of magnitude, respectively, below the 
production at RHIC. We note that the SPS data does not extend above $p_T > 2$ GeV/$c$, but 
the overall lower production of anti-protons makes it unlikely that anti-proton excesss 
above the fragmentation values is present.  

At RHIC energies, most of protons are produced in the reaction rather than transported 
from the beam. Both proton and anti-proton yields are enhanced with respect to the pion 
yields. Most often the observed enhancement is interpreted as a consequence of a strong 
radial flow which pushes the heavier particles to larger $p_{T}$. In this paper we address 
two outstanding questions:      
(1) whether the anomalous baryon enhancement at RHIC is a mass effect, or a baryon/meson effect. (2) what fraction of the baryons at intermediate $p_T$ originate in
the fragmentation. The PHENIX experiment measured $\phi$ meson production and jet 
correlations with identified particle trigger with the goal to solve the baryon puzzle at 
RHIC. We also include results from $\snn = $ 62.4 GeV  Au+Au collisions, which will help 
to understand the energy dependence of the observed phenomena.  

\begin{figure}[t]
\includegraphics[width=1.0\linewidth]{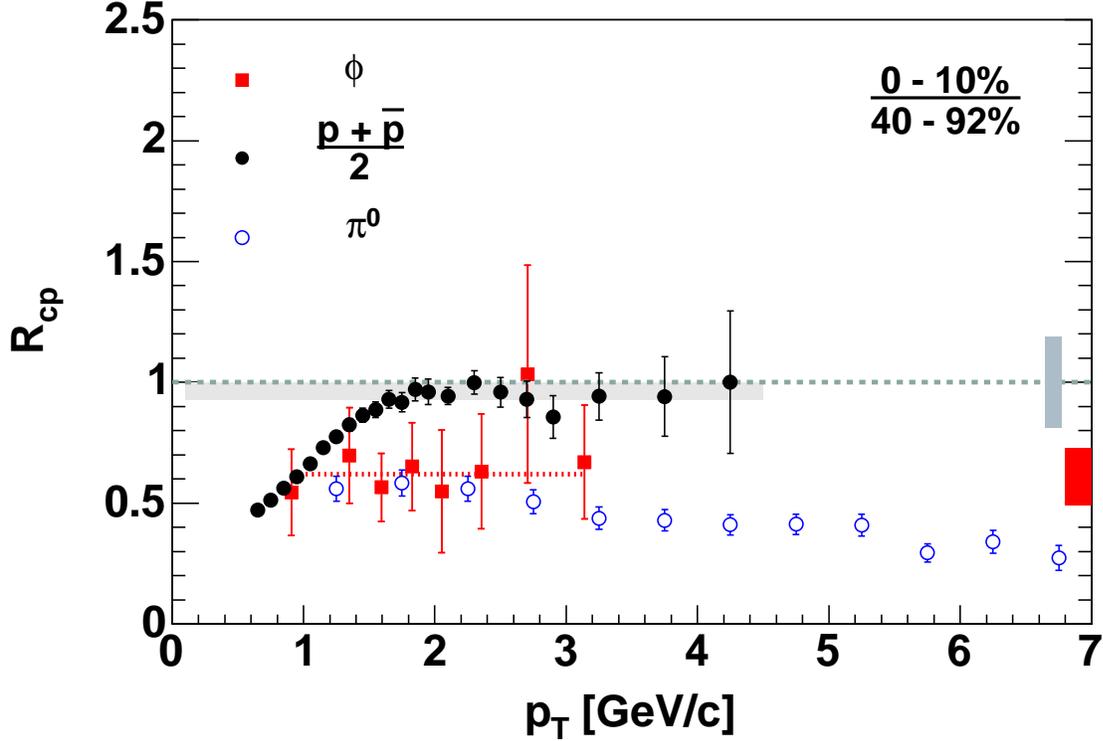}
\caption{The $R_{CP}$ of the $\phi$ mesons as measured in the $K^{+}K^{-}$
channel~\cite{ppg016,HQ04_DD}, compared to the protons, antiprotons, and $\pi^{0}$ 
for Au+Au collisions at $\snn = $ 200 GeV.}
\label{fig:phircp}
\end{figure}   

\section{Production of $\phi$ mesons}

The $\phi$ meson is an ideal test particle for investigating if the observed baryon
anomaly is cased by just the strong radial flow (mass effect) or not,
since $\phi$ is a heavy meson  that has mass similar to the mass of the  proton. 
Figure~\ref{fig:phircp} shows R$_{CP}$, the ratio of the yields in central to 
peripheral Au+Au collisions scaled by binary collisions, for protons, pions 
and $\phi$ mesons. The $\phi$ is  detected via its  $KK$ decay channel~\cite{qmproc,ppg016,HQ04_DD,ppg048}.
The $\phi$ follows the suppression pattern of the pions within errors, 
indicating that the observed behavior of the protons is not characterized
by their mass, but rather - by the number of valence quarks (meson or baryon). 
The clear baryon and meson effect at intermediate $p_T$ has also been observed by 
the STAR experiment in $R_{CP}$ of $K^{*}$, $K^{0}_{s}$, $\Lambda$, and 
$\Xi$~\cite{starLambda,starQM04Lamont}. 
In the hydrodynamical model with the jet fragmentation~\cite{hydro}, 
it is found that there is a general agreement in $p/\pi$ and $\bar{p}/\pi$ ratios 
between data and the model. But it might be challenging for the model to reproduce 
the suppression pattern on $\phi$ and $K^{*}$ by the hydrodynamical framework, 
since the heavier particle should gain a similar $p_T$ boost by the strong hadronic flow, 
regardless whether they are baryons or mesons. On the other hand, the recombination 
models~\cite{recoDuke,recoOregon,recoTAMU} give a more natural explanation on the 
baryon/meson effect, since the essential ingredient of this model is the number of 
 quarks in a hadron. 

\section{Jet correlations}

A crucial test of the origin for the baryon enhancement is to see if baryons 
in the intermediate $p_T$ exhibit a characteristic correlation of jets from 
hard-scattered partons. We have measured two particle correlations where 
the trigger particle is an identified meson ($\pi$, $K$) or baryon ($p$, $\bar{p})$ 
at 2.5 $< p_T< $  4.0 GeV/$c$~\cite{ppg033}. Associated particles, i.e. lower 
$p_T$ charged hadrons, near in azimuthal angle to the trigger are counted
(conditional yield per trigger). 

\begin{figure}[t]
\includegraphics[width=1.0\linewidth]{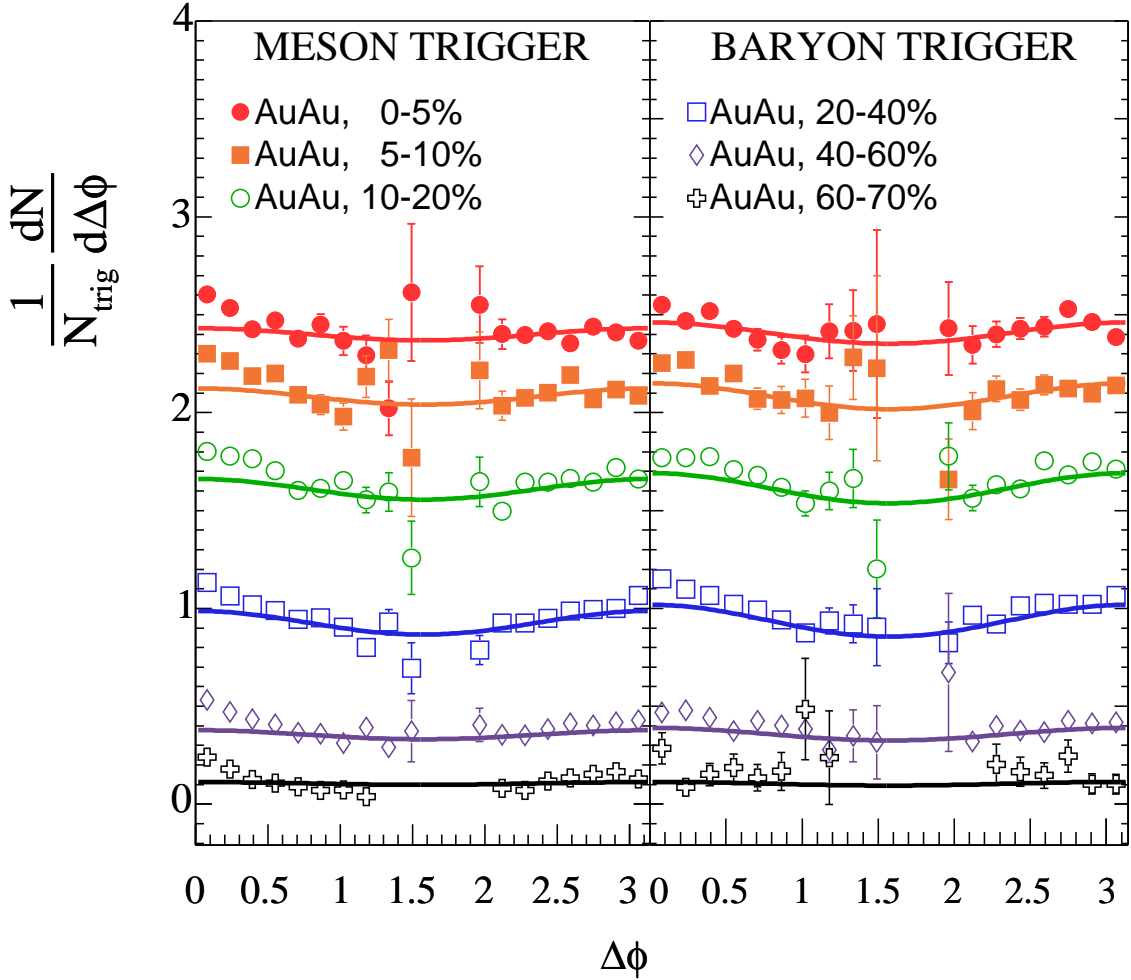}
\caption{$\Delta\phi$ distributions for meson (left) and baryon (right) 
triggers with $2.5 < p_T < 4.0$ GeV/$c$ and associated charged hadrons 
with $1.7 < p_T < 2.5$ GeV/$c$ for six centralities in Au+Au collisions~\cite{ppg033}. 
The solid lines indicate the calculated combinatorial background in the 
event modulated by the measured elliptic flow.}
\label{fig:jet_cor}
\end{figure}   

\begin{figure}[t]
\includegraphics[width=1.0\linewidth]{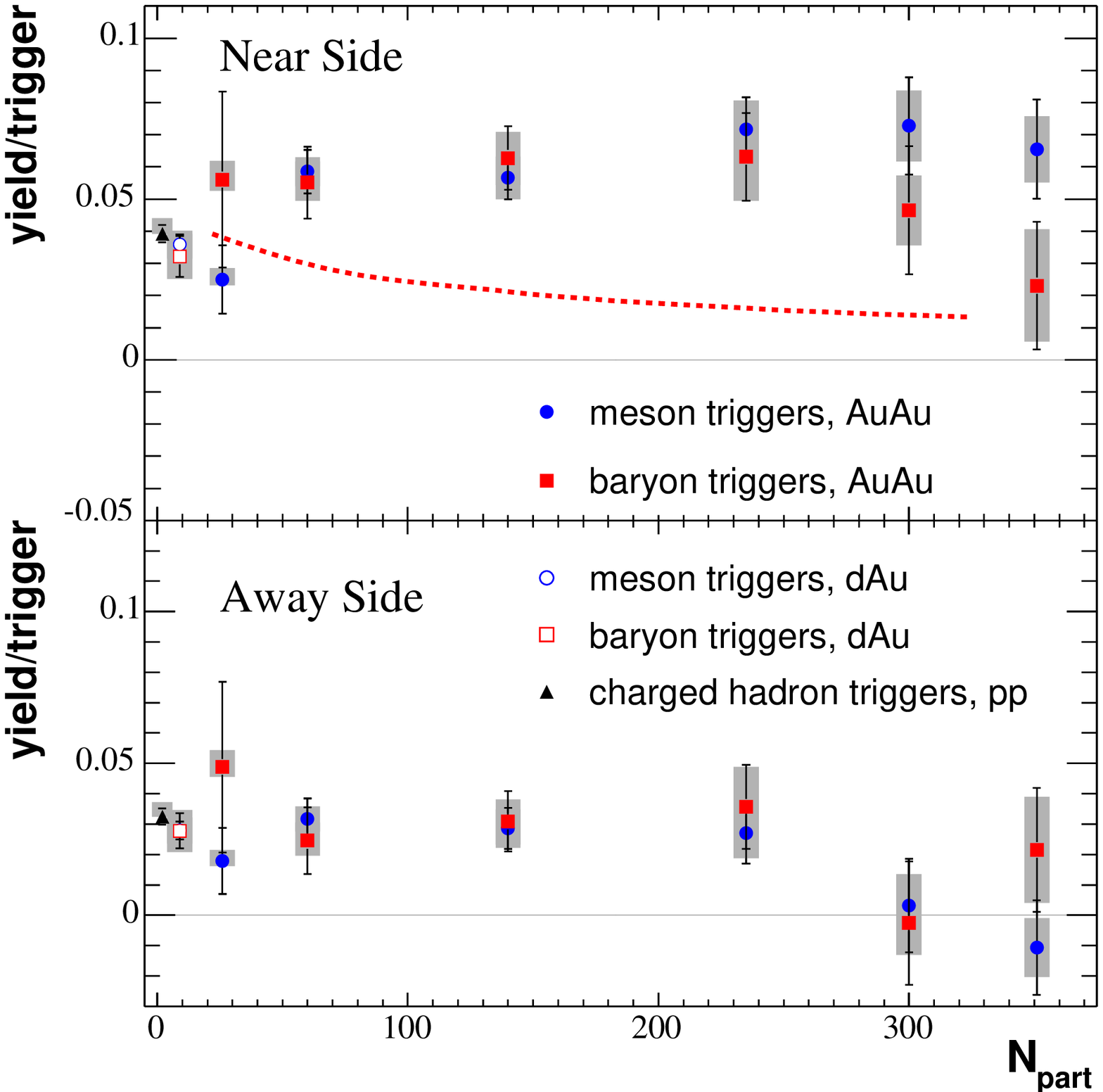}
\caption{Yield per trigger for associated charged hadrons between 
$1.7 < p_T < 2.5$ GeV/$c$ for the near- (top) and away- (bottom) 
side jets~\cite{ppg033}. The dashed line (top) represents an upper 
limit of the centrality dependence of the near-side partner yield 
from thermal recombination.}
\label{fig:jet_yield}
\end{figure}   

Figure~\ref{fig:jet_cor} shows the $\Delta\phi$ distributions with trigger mesons 
(left) and baryons (right) measured at midrapidity in Au+Au collisions at $\snn = 200$ GeV. 
The jet-like correlations are clearly seen over the combinatorial background, particularly 
at small relative angle inside the expected jet cone.
Figure~\ref{fig:jet_yield} shows the conditional yield per trigger of partner particles 
in p+p, d+Au, and Au+Au collisions, as a function of centrality. The top panel shows 
partner yield at small relative angle, from the same jet as the trigger hadron. 
We observe an increase in partner yields in mid-central Au+Au compared to the d+Au and 
p+p collisions. In Au+Au collisions, the near side yield per {\it meson} trigger remains 
constant as a function of centrality, whereas the near-side yield per {\it baryon} trigger 
decreases in the most central collisions as expected if a fraction of the baryons were 
produced  by soft processes such as recombination of thermal quarks. 
The dashed line in Figure~\ref{fig:jet_yield} represents an upper limit to the centrality 
dependence of the jet partner yield from thermal recombination. The data clearly disagree 
with both the centrality dependence and the absolute yields of this estimation, indicating 
that the baryon excess has the same jet-like origin as the mesons, except perhaps in the 
highest centrality bin. The bottom panel of Figure~\ref{fig:jet_yield} shows the 
conditional yield of partners on the away side. It drops equally for both trigger baryons 
and mesons going from p+p and d+Au to central Au+Au, in agreement with the observed 
disappearance \cite{starb2b} and/or broadening of the dijet azimuthal correlations. 
It further supports the conclusion that the baryons originate from the same jet-like 
mechanism as mesons.

\begin{figure}[t]
\includegraphics[width=1.0\linewidth]{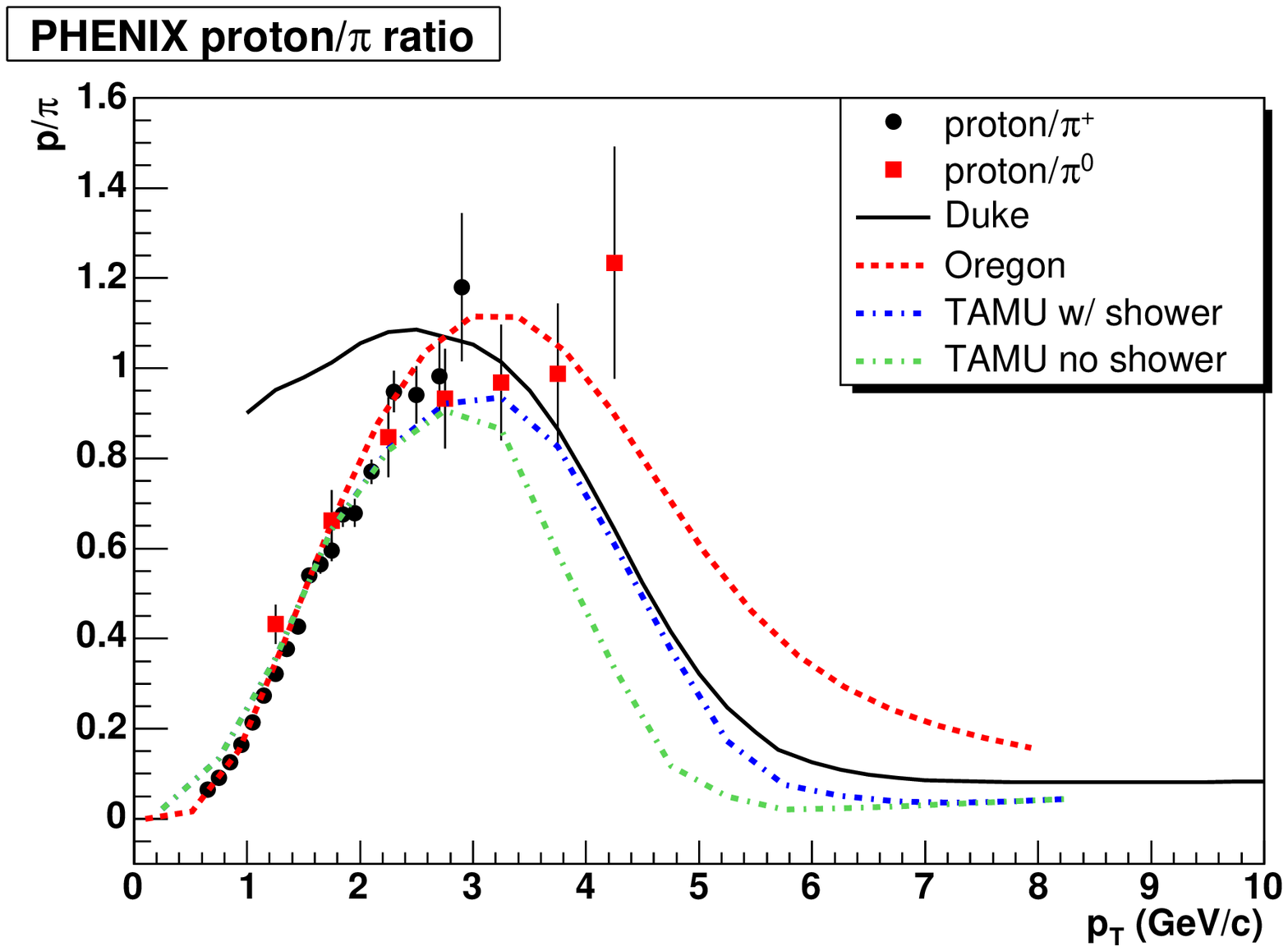}
\caption{The proton to pion ratio measured by PHENIX for Au+Au collisions 
at $\sqrt{s_{NN}}=200 $GeV with several comparisons to recombination models. 
The figure taken from~\cite{ppg048}.}
\label{fig:ppivsmodel}
\end{figure}

Recently several version of recombination models have been proposed, e.g. 
a purely thermalized quark recombination~\cite{recoDuke}, the model including 
shower partons or hard partons with thermal partons~\cite{recoOregon,recoTAMU}.
Figure \ref{fig:ppivsmodel} shows those recombination model calculations 
compared to the $p/\pi$ ratio from PHENIX (the figure taken from Ref.~\cite{ppg048}). 
All models are able to reproduce the general features of data at $p_T>3$ GeV/$c$. 
By including the hard-thermal component in the model, it seems there is 
a better agreement with the data, which is consistent with results of 
two particle jet correlations as shown above. It should be noted that 
the simple picture of recombination of purely thermal quarks~\cite{recoDuke} 
may apply, since the conditional yield with near-side baryons in most 
central collisions appears to decrease.

\section{p/$\pi$ and $\bar{p}/\pi$ ratios at $\sqrt{s_{NN}} = $ 62.4 GeV}

\begin{figure}[t]
\includegraphics[width=1.0\linewidth]{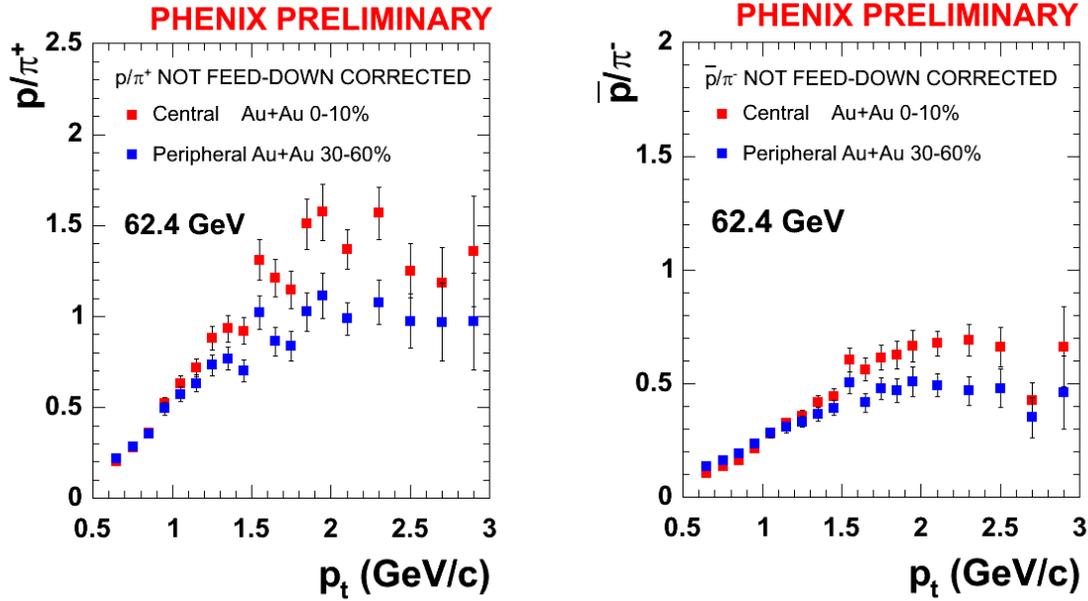}
\caption{$p/\pi^{+}$ (left) and $\bar{p}/\pi^{-}$ (right) ratios 
for central (0-10\%), and non-central (30-60\%) Au+Au collisons 
at $\snn = $ 62.4 GeV. Note that the feed-down corrections 
from weak decays are not corrected.}
\label{fig:ppi_62GeV}
\end{figure}

To study the excitation function of hadron production at the beam energy between 
CERN-SPS ($\snn = 17.3$ GeV) and RHIC, the data in $\snn = 62.4$ GeV in Au+Au 
was taken as a part of Run-4 RHIC program in 2004. This lower energy data 
provides an important information on the baryon production and transport 
at mid-rapidity between SPS and RHIC. Figure~\ref{fig:ppi_62GeV} shows 
the $p/\pi^{+}$ and $\bar{p}/\pi^{-}$ ratios for central (0--10\%), and 
non-central (30--60\%) Au+Au collisons at $\snn = $ 62.4 GeV. It should be
noted that the feed-down corrections from weak decays are not yet applied to these
results. The data shows a similar large proton contribution at intermediate $p_T$ 
as seen in 130/200 GeV Au+Au, while there is a less antiproton contribution 
compared to the 200 GeV results. Proton-antiproton pair production 
is reduced, and the baryon chemical potential is larger at 62.4 GeV than that 
of 200 GeV in Au+Au collisions. 
 
\section{Summary}
We presented the results on the production of $\phi$ mesons and two-particle 
azimuthal angular correlations with both the intermediate $p_T$ baryon and meson 
trigger in Au+Au collisions at RHIC. The $R_{CP}$ of $\phi$ follows the suppression 
pattern of $\pi^{0}$'s, which indicates that the baryon enhancement at intermediate 
$p_T$ in central Au+Au is not related to the paticle mass, but rather - to the number 
of valence quarks in the hadron. In the correlations study using identified 
particle trigger, the result shows that a large fraction of the baryons produced
at intermediate $p_T$ in Au+Au collisions originate from  jet fragmentation, 
perhaps except for the most central reactions. The recombination models which include 
hard-thermal component are in better agreement with the data than models that treat 
recombination of thermal quarks only.  The first measurements of $p/\pi^{+}$ and 
$\bar{p}/\pi^{-}$ ratios at $\snn = 62.4$ GeV in Au+Au collisions are studied.The data 
indicates a proton-antiproton pair production is reduced, and the baryon 
chemical potential is larger at 62.4 GeV than that of 200 GeV in Au+Au collisions.


\end{document}